\documentclass{aa}
\usepackage{graphics}
\begin{document}
 
   \thesaurus{11.07.1, 11.05.1, 11.09.4, 11.03.3, 13.09.1, 13.09.3}

   \title{ Dust in elliptical galaxies: a new dust mass 
evaluation}

\author{P. Merluzzi } 

 
   \institute{  Osservatorio Astronomico di Capodimonte, via Moiariello 16\\
 	      I-80131 Napoli, Italy\\
              email: merluzzi@na.astro.it}
 
   \date{Received 17 November 1997 / Accepted 5 August 1998}

   \authorrunning{P. Merluzzi}
   \titlerunning{Dust in elliptical galaxies}
 
   \maketitle
 
   \begin{abstract}
In order to investigate the nature and origin of dust in elliptical galaxies, a
method for the dust mass evaluation, which accounts for the dust temperature 
distribution, is here presented and discussed.
The derived dust masses turn out to be a 
factor 2-6 larger than those obtained with the single temperature
approximation. A correlation between the far-infrared and the blue 
luminosity has been also found.
The results are discussed in terms of dust ``mass discrepancy''
and of the possible evolution scenarios: {\it evaporation flow} and/or
{\it cooling flow}. While the present data cannot discriminate between these two 
scenarios, it is conceivable that the dust in elliptical galaxies can be 
accreted by the contribution of different mechanisms, according to the history 
of the individual objects.
\end{abstract}

\section{Introduction}\label{introduction}

Evidence for the presence of dust in 
elliptical galaxies was given by the optical observations of obscured regions 
in some systems (see for instance Bertola et al. 1985, V\'eron \& V\'eron 1988)
and was finally confirmed by the FIR observations of the IRAS satellite 
(Knapp et al. 1989, Roberts et al. 1991).
The low resolution IRAS data ($>$ 1 arcmin) have to be processed by 
adequate techniques to provide detailed ($\sim 1$ arcmin at 100 $\mu$m) 
information about the dust spatial distribution in nearby 
elliptical galaxies. In general, only the integrated flux of the detected 
source is available in the IRAS bands.
 
Despite the 
intrinsic limits due to the low resolution, Knapp et al. (1989) showed that a 
significant fraction (48$\%$) of the nearby 
E and S0 galaxies from the Revised Shapley-Ames Catalogue (Sandage \& Tammann
1981,
hereafter RSA) have been detected by IRAS at 60 and 100 $\mu$m at the limiting 
sensitivity (about 3 times lower than in the IRAS Point Source Catalog).
 
It is not surprising that ellipticals contain dust, since the presence
of dust is directly related to the stellar formation. But the 
coexistence of solid particles with the dominant gas component, which 
in these 
galaxies is heated to $\sim 10^7$ K and radiates at X-ray wavelengths, is a
matter of discussion. In fact, dust grains should be quickly destroyed by 
sputtering (Draine \& Salpeter 1979) when in direct contact with 
the hot gas. In such an environment the dust has a lifetime of $10^6-10^7$ yr.
The question thus follows: where does the dust come from?

At FIR wavelengths (60 and 100 $\mu$m) the thermal emission is mainly due to 
large grains
(radius $\sim 0.1$ $\mu$m) which may have different heating sources, e.g. 
the general interstellar radiation field or OB stars. In order to 
discriminate between the two 
different contributions and to estimate the weight of each of them, several 
efforts were undertaken (see for instance Calzetti et al. 1995).
A reasonable approach is to consider the 60 $\mu$m flux entirely due to
the warm dust ($\sim$40 K), while the 100 $\mu$m flux should be considered the
result of two contributions: the warm and the cold ($\sim$10 K) dust. While 
a two-component dust model is used to explain the spectral trend for 
different types of galaxies, it is rarely adopted for elliptical 
galaxies, which are often characterized by weak FIR emission and which
are not always detected in all IRAS bands. Therefore, a single 
color temperature (from the 60 and 100 $\mu$m data) is usually taken as the 
dust
temperature. It follows that no information about the dust temperature 
distribution and the dust spatial distribution is available for
elliptical galaxies. 
The
availability of the ISO data will provide spectra in a wider IR 
wavelength range (2.5-240 $\mu$m). Several attempts to understand the dust 
nature and origin suggest interesting interpretations by comparing optical
and FIR data (Goudfrooij \& de Jong 1995, hereafter GJ95, Tsai \& Mathews 1995,
1996), by studying the stellar content, or by using a severe and critical 
approach to the data (Bregman et al. 1998). Finally, few elliptical galaxies 
have been observed at sub-millimeter wavelengths by Fich \& Hodge (1991, 1993).

GJ95 found that the dust masses determined 
from the  
IRAS flux densities are rou\-gh\-ly an order of magnitude higher than those 
determined from optical extinction values of dust lanes and patches, in 
contrast with what happens for spiral galaxies. The authors suggest that this 
``mass discrepancy'' may be explained by the existence of a diffuse component
(within 2 Kpc from the center), 
which is not detectable at optical wavelengths.
On the other hand, Tsai \& Mathews (1996) suggested that, while the 
distributed dust component is associated with dust recently ejected from 
evolving stars, another ``extra'' component of dust is present in ellipticals
both in dust lanes and rings and/or in other galactic regions. In
particular, they postulated that a substantial mass of cold gas remains
``observationally elusive without forming completely into stars''.
If the extra dust is 
optically thin in the visible it
should be located far from the galactic core region, where the intensity of 
the starlight and therefore the grain heating is reduced.

The dust spatial 
distribution suggested by GJ95 and by Tsai \& Mathews 
(1996) support the two most popular scenarios 
respectively: the {\it evaporation flow} picture and the {\it cooling flow} 
picture. In the {\it evaporation flow} scenario the clouds of dust and gas 
currently observed
in ellipticals have mainly an external origin, being associated to events of
galaxy interaction and/or mergers and being heated by thermal conduction in 
the hot gas (de Jong et al. 1990, Sparks et al. 1989). 
On the contrary ({\it cooling flow}), 
the internal origin of gas and dust may be explained with both red giant winds 
(Knapp et al. 1992) and by the {\it cooling flow} mechanism, in which mass loss 
from stars within the galaxy is heated by supernova 
explosions and by collisions between expanding stellar envelopes during the 
galaxy formation stage, and then cools and condenses (Fabian et al.1991).

Therefore, in ellipticals, the presence of dust and the dust ``mass 
discrepancy'' are related with the dust spatial distribution,
which depends on the nature and the evolution of these systems.
Hence, in order to investigate the dust content, dust mass 
evaluations as accurate as possible are required to estimate
the amount of the ``mass discrepancy''.
Since both scenarios suggest the presence of different dust components
and the very nature of the galactic environments involves the existence of
dust grains at different temperatures, it is necessary to take into account 
a dust temperature distribution for the dust mass evaluations.

Kwan $\&$ Xie (1992) suggest a theoretical approach to take into account the
effects of the dust temperature distribution in the dust mass evaluation.
I present here an application of their method which
is discussed and implemented in Section 2. The results obtained for the
sample of ellipticals, introduced in Section 3, are presented and 
discussed in Section 4 by comparing the 
FIR with the visual dust mass evaluation and by discussing the
correlation between blue and FIR luminosities.

\section{Dust mass evaluation from FIR observations}

The dust mass can be derived both from optical and from FIR observations.
The value of the mass depends on the physical-chemical properties of the
solid particles (i.e. grain radius, grain density and emissivity).
The usual approach consists of assuming some values for the grain properties
(Hildebrand 1983), while a color temperature is derived from the FIR emission.
Different authors, however, present slightly different formulae for the 
evaluation of the dust mass
(Thronson \& Telesco 1986, Greenhouse et al. 1988, Young et al. 1989, Roberts 
et al. 1991 and Thuan \& Sauvage 1992). The differences between these 
relations are only due to
different assumptions on the grain parameters and to different derivations 
of the color temperature.
It should be noticed that, within a flux uncertainty of 10$\%$ (which is 
typical of the high quality IRAS data), all the dust mass values obtained by 
using the different formulae are in agreement.
 
A single temperature 
model is a rough approximation in describing a galactic environment.
The dust is in fact heated by the radiation
field, which in turn depends on the sources of luminosity and on their spatial
distribution in the galaxy. The total FIR emission is thus likely due to the
contribution of dust at different temperatures.  Moreover, the IRAS FIR
measurements are not adequate to detect the emission coming from cold dust
($10-20$ K) which peaks at wavelengths between 200 and 300 $\mu$m.

I adopt the dust temperature distribution given by Kwan \& Xie (1992):
\begin{equation}
g(T) = {{(T/T_L)^\gamma e^{-\beta(T/T_L)}}\over{{\int_{T_L}^{T_U} 
(T/T_L)^\gamma e^{-\beta(T/T_L)}dT}}}\hspace{0.2cm}for~~T_L \leq T \leq T_U\ ,\
\label{eq:curv}
\end{equation}
\begin{equation}
g(T) = 0 \hspace{3.8cm}otherwise.
\end{equation}
$T_L$ and $T_U$ are the lower and upper limits of the temperature $T$; 
$\beta$ and $\gamma$ are free parameters that determine the shape of
the distribution. 
The equations relating the temperature distribution $g(T)$ to the luminosity
emitted by the dust and to the dust mass are detailed in Kwan \& Xie (1992).

Due to the observed spectral range, I adopt $T_L$=7 K and $T_U$=60 K, taking
into account only temperature distributions peaking at a value intermediate
between $T_L$ and $T_U$ and excluding those pairs of $\beta$ and $\gamma$ 
which produce unrealistic distributions (as, for instance, monotonically
increasing or decreasing functions).

The main problem in the present approach is to select the proper values for 
the parameters $\beta$ and $\gamma$, the choice being constrained by the ratio 
of the flux densities at two different wavelengths.

I first identify a range of
$\beta$ and $\gamma$ pairs which produce the observed flux ratio.
Since the same flux ratio can be obtained by functions having quite 
different shapes, a further constraint is needed.
Unfortunately, for none of the galaxies in the sample submillimeter 
observations are available. 
Nevertheless, their color temperature may be derived from the flux density 
ratio (Henning et al. 1990).  
Taking into account the uncertainty in the computed color temperature, I
select those distributions $g(T)$ that produce the observed flux density
ratio and whose peak temperature is comparable to the color temperature.
The selected family of $g(T)$ functions obviously satisfies
two conditions which are not independent (being both related to the flux 
ratio). This fact could in principle affect the reliability of the
results. It turns out, however, that pairs of $\beta$ and $\gamma$ which 
satisfy the same constraints
give the same dust mass within the flux uncertainty.

In order to check the method a simple numerical simulation has been performed.
I considered an artificial galaxy which is not resolved by IRAS both at 60 and 
100 $\mu$m with a given dust mass and, then, I evaluated the dust content
following the present method. The galaxy dust mass $M_d$ may be roughly 
estimated by using the equation
\begin{equation}
M_d = {{4}\over{3}} \pi r^3 \rho N_d\ ,
\label{eq:mass}
\end{equation}
where $r$ is the dust grain radius, $\rho$ is the dust grain density and
$N_d$ is the number of dust grains. 
Eq. 3 implies the following approximations. The dust grain radius $r$ 
characterizes the dust grains which contribute to the thermal emission
at FIR wavelengths. Actually, the uncertainties in the total dust
mass introduced by using an average radius instead of a size 
distribution, are much smaller than those arising from uncertainties 
in the dust emissivity spectral trend and in the dust temperature distribution
(Kwan $\&$ Xie 1992).
$N_d$ is the total number of the dust grains within the observing beam. Since
the galaxy is not resolved by IRAS, as it is often the case, Eq. 3
gives the total dust mass of the point source. A dust radius of 0.1 $\mu$m and 
a dust grain density of 3 g cm$^{-3}$ are currently adopted (Hildebrand 1983) 
and also used in the present article. 

Concerning the dust emissivity, the power-law 
approxi\-mation $\epsilon(\lambda) = \epsilon_0(\lambda_0/\lambda)^\alpha$
is used in the computations, with $\alpha = 1$ and $\epsilon_0 = 9.38\cdot
10^{-4}$ (Hildebrand 1983).

By assuming a value for $N_d$ and the 
temperature distribution $g(T)$, and accounting for the spectral responses
of the IRAS detectors at 60 and 100 $\mu$m, the luminosity emitted by the 
$N_d$ dust grains, as observed within the filter bandpass, 
can be estimated. The flux ratio 
and the color temperature of the galaxy are then computed, while 
the dust mass of the source is derived from $N_d$ by Eq. 3.

A galaxy with $M_d$=$10^5$ M$_\odot$ turns out to contain about $N_d$=1.6$\cdot 
10^{52}$ dust grains. The galaxy distance is assumed to be 10 Mpc.
By assuming a $g(T)$ with $\beta=5$ and $\gamma=20$ the flux ratio turns out to 
be about 0.5. I 
use the flux ratio as a constraint to select the pairs $\beta$-$\gamma$ 
producing a $g(T)$ whose peak 
temperature is comparable to the galaxy color temperature of 36 K.
Among these pairs also the pair $\beta=5$ and $\gamma=20$ is found (i.e. 
exactly the values adopted for the input $g(T)$).
For the selected pairs of parameters I compute the dust mass which 
ranges between $6-9\cdot10^4$ M$_\odot$. Therefore, the derived values are in 
agreement with the dust mass of the galaxy within the flux uncertainties,
thus confirming the reliability of the method. 
Futhermore, by using the single temperature
model and taking into account the different formulae available
(see sect. 2) a dust mass of $4\cdot10^4$ M$_\odot$ is derived.
This result shows that the single temperature model 
underestimates the dust content. 

The same test was performed for different ``artificial galaxies'' with different
tem\-pe\-ra\-tu\-re distributions obtaining always consistent results.

\section{The sample of elliptical galaxies}

\begin{table}

\caption[]{The Selected Sample of Elliptical Galaxies.}

\[
\begin{array}{rrcccccc}
\hline
\noalign{\smallskip}
{\rm Galaxy} & {\rm log L_B} & {\rm log L_{FIR}} & {\rm {f_{60}}\over{f_{100}}}
& {\rm T_{dust}} & {\rm M^{opt}} & {\rm M_{T}^{FIR}} &
{\rm M_{g(T)}^{FIR}}\\
& {\rm [L_\odot]} & {\rm [L_\odot]} & & {\rm [K]} & {\rm [M_\odot]}
& {\rm [M_\odot]} & {\rm [M_\odot]} \\
{\rm (1)} & {\rm (2)} & {\rm (3)} & {\rm (4)} & {\rm (5)} & {\rm (6)} &
{\rm (7)} & {\rm (8)}\\
\noalign{\smallskip}
\hline
\noalign{\smallskip}
{\rm N1395} & 10.36 & 8.27 & 0.17 & 25.0 &               & 5.43\cdot 10^5
& 1.61\cdot 10^6 \\
{\rm N2974} & 10.29 & 9.16 & 0.25 & 28.1 & 4.57\cdot10^4 & 1.77\cdot 10^6
& 1.11\cdot 10^7 \\
{\rm N3377} &  9.70 & 7.56 & 0.45 & 33.8 & 9.77\cdot10^3 & 1.39\cdot 10^4
& 5.06\cdot 10^4 \\
{\rm N3557} & 10.86 & 9.22 & 0.37 & 31.7 &               & 9.42\cdot 10^5
& 3.16\cdot 10^6 \\
{\rm N3904} & 10.16 & 8.45 & 0.44 & 33.4 &               & 1.15\cdot 10^5
& 5.01\cdot 10^5 \\
{\rm N4125} & 10.75 & 9.08 & 0.49 & 34.7 & 4.47\cdot10^5 & 3.90\cdot 10^5
& 1.46\cdot 10^6 \\
{\rm N4261} & 10.56 & 8.35 & 0.62 & 38.0 &               & 4.47\cdot 10^4
& 1.49\cdot 10^5 \\
{\rm N4278} &  9.76 & 8.25 & 0.36 & 31.4 & 2.34\cdot10^4 & 1.05\cdot 10^5
& 4.75\cdot 10^5 \\
{\rm N4374} & 10.43 & 8.43 & 0.50 & 35.0 & 3.47\cdot10^4 & 8.38\cdot 10^4
& 3.00\cdot 10^5 \\
{\rm I\,3370} & 10.69 & 9.59 & 0.28 & 29.1 & 3.47\cdot10^5 & 3.87\cdot 10^6
& 2.11\cdot 10^7 \\
{\rm N4486} & 10.67 & 8.45 & 1.11 & 49.5 & 1.48\cdot10^3 & 1.52\cdot 10^4
& 3.79\cdot 10^4 \\
{\rm N4589} & 10.29 & 8.92 & 0.36 & 31.2 & 8.51\cdot10^4 & 5.10\cdot 10^5
& 1.44\cdot 10^6 \\
{\rm N4697} & 10.46 & 8.59 & 0.43 & 33.2 &               & 1.64\cdot 10^5
& 7.31\cdot 10^5 \\
{\rm N4696} & 10.83 & 9.06 & 0.14 & 23.7 & 4.47\cdot10^5 & 5.19\cdot 10^6
& 3.02\cdot 10^7 \\
{\rm N5018} & 10.63 & 9.67 & 0.59 & 37.4 & 2.82\cdot10^5 & 9.90\cdot 10^5
& 3.38\cdot 10^6 \\
{\rm N5044} & 10.54 & 8.68 & 1.08 & 48.8 & 1.82\cdot10^4 & 2.76\cdot 10^4
& 6.76\cdot 10^4 \\
{\rm I\,4296} & 11.00 & 9.02 & 0.61 & 37.8 &               & 2.13\cdot 10^5
& 7.00\cdot 10^5 \\
{\rm N5322} & 10.77 & 9.07 & 0.48 & 34.6 &               & 3.89\cdot 10^5
& 1.37\cdot 10^6 \\
{\rm N5576} & 10.27 & 8.13 & 0.47 & 34.4 & 3.31\cdot10^3 & 4.59\cdot 10^4
& 1.47\cdot 10^5 \\
{\rm N7144} & 10.30 & 8.54 & 0.31 & 29.9 &               & 2.83\cdot 10^5
& 1.16\cdot 10^6 \\
{\rm I\,1459} & 10.54 & 9.00 & 0.50 & 35.0 & 1.86\cdot10^5 & 3.13\cdot 10^5
& 2.09\cdot 10^6 \\
\noalign{\smallskip}
\hline
\end{array}
\]
Column (2): total blue luminosities taken from
GJ95. Columns (3), (4) and (5): computed
values for FIR luminosities, flux ratio and dust color temperatures.
The dust masses in Column (6) are taken from GJ95.
The FIR dust mass
in Column (7) is derived by averaging the results of formulae which are
usually adopted, while the FIR dust mass in Column (8) is derived by adopting
a temperature distribution model as described in Sect. 2. The
uncertainties for the mass evaluations are shown in Fig. 1. N and I stand
for NGC and IC respectively.
\end{table}

The 21 elliptical galaxies listed in Table 1 have been extracted from the sample
by GJ95, who evaluated their FIR luminosity and dust
mass from FIR data. All the galaxies are classified as E both in RSA and in 
de Vaucouleurs et al. (1991) and have magnitudes $B^0_T < 12$. I selected those 
galaxies which have been detected at both
60 and 100 $\mu$m by IRAS (Knapp et al. 1989). H$_0$ = 50 Km s$^{-1}$Mpc$^{-1}$
is assumed throughout this paper.

For each galaxy the FIR luminosity, the flux ratio, the dust temperature and 
the dust mass from FIR data (hereafter FIR dust mass) have been derived and 
listed in Table 1. The FIR fluxes by Knapp et al. (1989) have been corrected
taking into account the contribution of hot circumstellar dust 
(GJ95) and have been used for the computation of:
the FIR luminosity in the band 40-120 $\mu$m (Helou et al. 1985), the 
flux ratio and the dust color temperature (Henning et al. 1990). The 
uncertainties in the three quantities depend on the flux 
uncertainties, which are estimated to be in the range 10$\%$-30$\%$ 
(Knapp et al. 1989). 
The chemical and physical 
properties of the dust grains play a critical r\^ole in the evaluation of 
the uncertainties on temperature and mass. On the other hand, the present 
paper is mainly addressed to the dependence of the dust mass on the choice 
of a particular temperature distribution, and I will not discuss
the problem of the dust properties (which will be studied in a forthcoming 
paper). A spectral index $\alpha=1$ for the dust is 
used in all the computations.
The FIR dust masses listed in Table 1 
are evaluated from the FIR fluxes by means of the single temperature and the 
temperature distribution models respectively.

\section{Temperature distribution model: results and discussion}

\subsection{FIR dust masses}

The dust masses computed by adopting a single temperature 
and a temperature distribution are compared in Fig.1.
Both methods lead to a temperature dependence, as 
expected when the dust amount is derived by using
thermal emission. In fact, the colder the dust grains, the larger 
should be their total number in order to produce a given FIR emission.

\begin{figure}
\resizebox{\hsize}{!}{\includegraphics{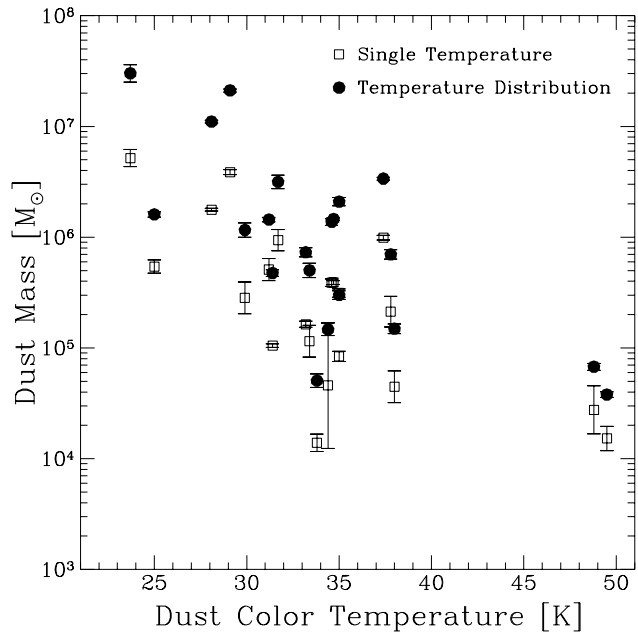}}
\caption{ The dust masses computed by adopting a single
temperature (open squares) and a temperature distribution (filled circles).
The error bars account for the uncertainties in $g(T)$ (filled circles)
and in the color temperature (open squares).}

\label{fp1}
\end{figure}

The dust masses obtained with a temperature distribution 
are larger than those derived from a single-tem\-pe\-ra\-tu\-re model.
This confirms that IRAS measurements allow to determine the warm (30-54 K) 
dust amount only, while the contribution of the cold dust is neglected.
Moreover, it turns out that, in the present sample, the FIR luminosity does 
not depend on the color 
temperature, thus confirming that the FIR emission is a result of different
contributions and cannot be properly explained by a source with 
a single equilibrium temperature.
The two models give masses differing by factors from 2 to 6.
This wide range is due either to the shape of the  
temperature distribution and/or to the uncertainties in the dust parameters. 
On the other hand, the ratio between the two dust mass 
evaluations (Fig. 2) shows a general temperature dependence which suggests 
that colder galaxies have a larger amount of missed dust.

\begin{figure}
\resizebox{\hsize}{!}{\includegraphics{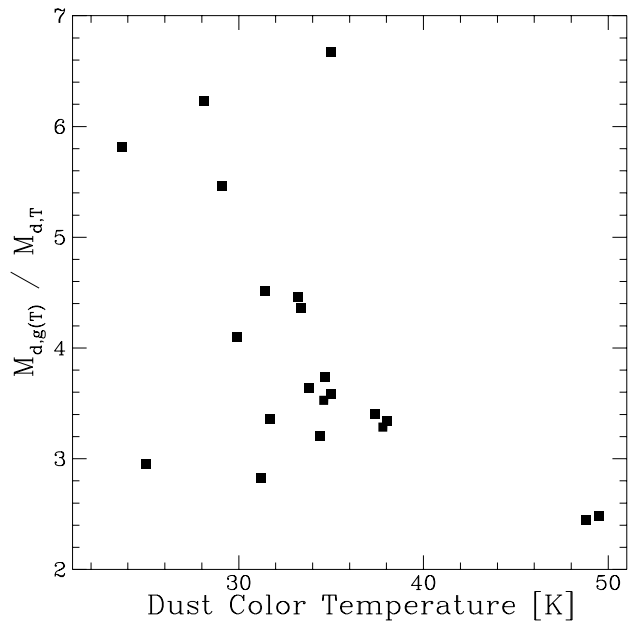}}
\caption{ The ratio between the FIR dust masses computed by using 
a temperature distribution and a single temperature model. The temperature 
dependence suggests that larger amounts of cold dust are neglected in sources 
with lower color temperature. The two galaxies on the left bottom, which do not
follow the general trend are, from left to right, NGC\,1395 and NGC\,4589.}

\label{fp2}
\end{figure}

\subsection{Dust ``mass discrepancy''}
 
The temperature distribution  model enhances the dust 
``mass discrepancy''. In order to explain that discrepancy,
it has to be noticed that the dust mass evaluated from optical observations
critically depends on the spatial distribution of the
dust with respect to the stars. Usually the 
extinction is thought to be due to an overlying absorbing screen of dust
grains. That is the geometry where a given amount of dust has the 
strongest effect on the starlight. In order to test how a peculiar and 
oversimplified spatial distribution affects the dust mass evaluation
from optical data, I compared the optical dust masses in Table 1 to those 
derived by assuming a more realistic spatial distribution (e.g. Witt et al. 
1992) and by using the optical data by Goudfrooij et al. (1994a, b). The new 
optical masses turn out to be enhanced by a factor 2-4, cutting the 
``mass discrepancy'' down.
However, even if the ``mass discrepancy'' is 
reduced by assuming a more realistic spatial distribution, the optical 
absorption can only account for the dust in the obscured galaxy regions,
while the distributed dust component is always neglected.

A similar case concerns the ``extra'' dust component suggested by
Tsai $\&$ Mathews (1996) to explain the 60-100 $\mu$m flux ratio:
due to its low temperature, it is not detected by IRAS. Following their 
model, it is possible to estimate the
contribution of the ``extra'' dust in the sample. Taking into 
account the ``extra'' dust amount, the dust masses derived by a 
single temperature model are slightly enhanced, but the large uncertainties
due to the observations and, then, to the temperature evaluation do not
allow to judge this enhancement as significative.
One can argue that, with a model of temperature distribution, it is possible to 
overestimate the dust amount. Actually, this cannot happen because of the 
severe constraints which are chosen, whether two different flux ratios are
available or when a flux ratio and the relative color temperature are used
as it is here suggested. In this respect, the last method is also the more 
conservative,
since the warm dust observed at 60 $\mu$m affects the dust mass evaluation
in the sense of underestimating the cold dust amount.

Therefore, since these computations are very sensitive 
to the assumed
dust tem\-pe\-ra\-tu\-re, the model by Kwan $\&$ Xie (1992) greatly improves 
the mass
evaluation by taking into account a wide tem\-pe\-ra\-tu\-re range, and 
can be applied to a wide range of galaxy morphological types when, because of 
lack of the data, it is no possible to adopt and to develop a suitable
radiation model. 

\begin{figure}
\resizebox{\hsize}{!}{\includegraphics{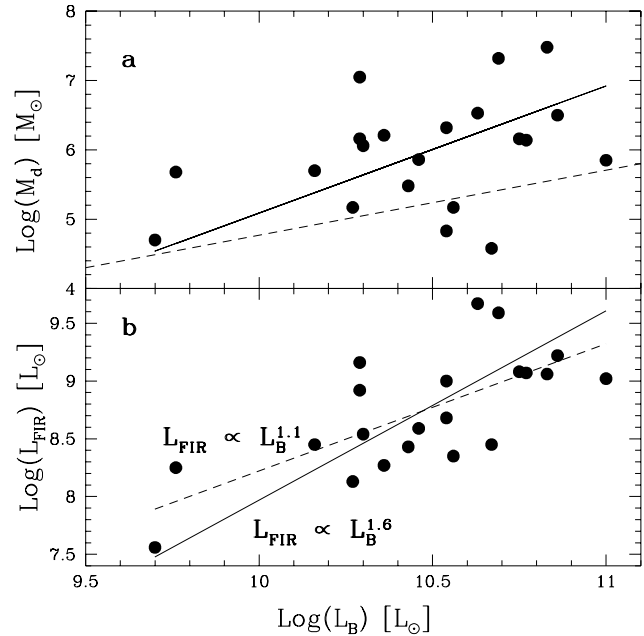}}
\caption{ {\bf a}) Dust mass (with the temperature distribution model)
and blue luminosity relation: the solid line is the
least squares bisector fit (confidence level greater than 95$\%$) 
while the dashed line is the loci
where dust is replenished by stellar mass loss and destroyed by sputtering 
with the maximum destruction timescale (10$^{7.5}$ yr). Most of the galaxies 
present an observed dust mass larger than the amount predicted by
the dust formation and destruction model.
{\bf b}) FIR and blue luminosity relation: the solid line is the 
least squares bisector fit (confidence level greater than 99.5$\%$).
For comparison the theoretical prediction (dashed line) by Tsai $\&$ Mathews
(1996) is also shown.}

\label{fp3}
\end{figure}

\section{Temperature distribution and heating mechanisms: the origin of the
dust}

Actually, the uncertainty on the temperature distribution is 
directly connected to the study of the heating mechanisms and, therefore, to 
the knowledge of the UV interstellar radiation field.

To this aim, and in order to investigate the origin of the dust in elliptical
galaxies, I plot in Fig. 3 a and b the dust mass M$_d$ and the FIR luminosity 
L$_{FIR}$ versus the blue luminosity L$_B$ for the galaxy sample. 
Both M$_d$ and L$_{FIR}$ are correlated to L$_B$, in contrast with what has 
been found by GJ95. Therefore, while the absence of correlation between dust 
mass and blue luminosity in their sample was considered to support
the external origin of the dust in ellipticals and the {\it evaporation flow}
picture, the present result indicates that a relationship between the dust
content and the present day population of stars should be not excluded.
But, as already pointed out by GJ95, a lack or a presence of correlation 
between M$_d$ and/or L$_{FIR}$ and L$_B$ cannot be a definitive proof of the 
external or the internal origin of the dust, due to the fact 
that the dust destruction mechanisms and timescales in elliptical galaxies are 
expected to be different depending on the evolutionary state and on the hot 
gas content of the individual objects.

For this reason, further analysis and, above all, high resolution FIR 
observations are required to understand the origin and the fate of dust in
these systems. Waiting for a complete analysis of the ISO data, a reasonable 
approach is the comparison between the available observations and models 
of star formation and dust heating mechanisms. I plot in Fig. 3a the loci
where dust is replenished by stellar mass loss at the rate given by
Faber $\&$ Gallagher (1976) and destroyed by sputtering with the maximum 
destruction timescale (10$^{7.5}$ yr) in ellipticals which contain hot gas (see
GJ95 for details). In 85$\%$ of the galaxies the dust having an internal 
origin does not account for the computed dust masses, which are larger. 
Therefore, one can argue that in 85$\%$ of the ellipticals an 
alternative supplying mechanism for the dust is required to account for the 
FIR observations. Mergers with spirals or small, dust-rich irregular galaxies
and , then, the {\it evaporation flow} scenario are, in fact, strongly supported by
the previous evidence (Fig. 3a). The critical point of 
this picture is to identify the observed diffuse dust, located within 2 Kpc
from the center, as the dust accreted during the galaxy interaction when
high hydrodynamic instabilities are expected. In particular, the external dust
had to be somehow protected from the interaction 
with the hot gas when moving toward the center.

The problem could be solved with the ``extra'' dust component (Tsai $\&$ 
Mathews 1996) located in very large disks out to the effective radius
and, therefore, cold enough to emit at FIR and submillimeter wavelengths.
This dust can have, in principle, both internal and external origin. In 
particular, due to the low temperature ``dust may re-form and grow in these 
cold disks'' (Tsai $\&$ Mathews 1996). Unfortunately, due to the large 
uncertainties, it is not possible to evaluate the exact amount of this
dust component. Therefore, since the external origin of the dust
is far from fully verified, I experimented with an alternative approach
to interpret the relation between the FIR and the blue luminosity.
By using the least squares bisector method, I find 
L$_{FIR}\propto $L$_{B}^{1.64\pm0.39}$ (Fig. 3b), in agreement with Bregman 
et al. (1998). 
The luminosity correlation supports the scenarios in which a 
significative amount of dust has internal origin (Tsai $\&$ Mathews 1995, 
1996), coming from stellar mass loss and being heated by stellar photons and the
general interstellar radiation field.

The problem of the coexistence of dust and hot gas might be resolved with a 
dust grain distribution containing grains larger than the maximum size 
suggested by Mathis et al. (1977) and usually adopted in the current  
models. In fact, this assumption would increase the sputtering time and, 
then, the dust grain density. Tsai $\&$ Mathews (1996) found that the FIR
luminosity is proportional to L$_{B}^{1.1}$ when the maximum grain size
increase from 0.3 $\mu$m to 0.9 $\mu$m. This 
theoretical prediction (dashed line in Fig. 3b) and the observed relation 
(solid line in Fig. 3b) is consistent in the luminosity range
covered by the present sample\footnote{Incidentally, an ordinary least squares 
fit to the data gives L$_{FIR}\propto $L$_{B}^{1.07\pm0.25}$.}, therefore the 
luminosity correlation seems to support the {\it cooling flow} scenario.

Finally, the two comparisons between observations and empirical-theoretical 
models show a clear conflict: while the canonical star formation rate cannot
account for the dust amount in elliptical galaxies, thus supporting the
{\it evaporation flow} scenario, the trend of the L$_{FIR}$-L$_B$ can be 
explained with a proper dust model which increases the sputtering time and
which suggests an {\it ad hoc} dust distribution. Concerning this last 
possibility it has to be stressed that, although the L$_{FIR}$-L$_B$ relation
can be affected by different hot gas contents and stellar populations 
in the individual objects, the correlation between these quantities
is characterized by a trend predicted by theoretical models. Furthermore,
both in Bregman et al. (1998) and in the present work, 
severe selection criteria are applied (see Sect. 3), which take into account 
the reliability of the observations and the different contributions to the FIR 
emission. I compare the present sample and the sample of 
Bregman et al. (1998) which rejected the galaxies whose FIR fluxes can be
contaminated by AGN emission, position uncertainties, background objects and 
inhomogeneity. The L$_{FIR}$-L$_B$ relation in Fig. 3b is not affected by the
exclusion from the sample of the four AGNs (NGC\,4374, NGC\,4486, I\,4296 and 
I\'1459) and of NGC\,3557 that has a non-homogenous background.

Taking into account the results of the present analysis it seems difficult to
choose between the {\it evaporation flow} or the {\it cooling 
flow} scenario and to affirm that the dust in ellipticals has mainly en external
or an internal origin. A resonable conclusion is that the individual objects
have experienced different mechanisms of dust accretion during their 
evolution and that only a detailed study of the individual systems will allow 
to describe for each object the different evolution phases and the 
contributions of the different mechanisms of dust accretion.

\acknowledgements{I thank M. Capaccioli for helpful discussions and for having
encouraged this research and the referee whose comments greatly helped in 
improving this work. I am also grateful to J.N. Bregman who has shown me 
his work on elliptical galaxies. }

----------------------------------


\begin{thebibliography}{}

\bibitem{} Bertola F., Galletta G., Zeilinger W., 1985, ApJ 292, L51

\bibitem{} Bregman J.N., Snider B.A., Grego L., Cox C.V., 1997, 
astr-ph/9802105, 9 Feb 1998 
 
\bibitem{} Calzetti D., Bohlin R.C., Kinney A.L., 1995, ApJ 44, 136

\bibitem{} de Jong T., Norgaard-Nielsen H.U., Hansen L., Jorgensen H.E., 
1990, A\&A 232, 317

\bibitem{} de Vaucouleurs G., de Vaucouleurs A., Corwin H.G., Buta R.G., Paturel
G., Fouqu\'e P., 1991, ``Third Reference Catalogue of Bright Galaxies'',
(Springer, New York)

\bibitem{} Draine B.T., Salpeter E., 1979, ApJ 231, 77

\bibitem{} Faber S.M., Gallagher J.S., 1976, ApJ 204, 365

\bibitem{} Fabian A.C., Nulsen P.E.J., Canizares C.R., 1991, A\&AR 2, 191

\bibitem{} Fich M., Hodge P., 1991, ApJ 374, L17

\bibitem{} Fich M., Hodge P., 1993, ApJ 415, 75

\bibitem{} Goudfrooij P., de Jong T., 1995, A\&A 298, 784, (GJ95)

\bibitem{} Goudfrooij P., de Jong T., Hansen L., Noergaard-Nielsen H.U.,
1994a, MNRAS 271, 833

\bibitem{} Goudfrooij P., Hansen L., J\o rgensen H.E., N\o rgaard-Nielsen H.U., 
1994b, A\&ASS 105, 341

\bibitem{} Greenhouse M.A., Hayward T.L., Thronson H.A., 1988, IAU Symp. No.
131, (Reidel:Dordrecht), p.170

\bibitem{} Helou G., Soifer B.T., Rowan-Robinson M., 1985, ApJ 298, L7

\bibitem{} Henning Th., Pfau W., Altenhoff W.J., 1990, A\&A 227, 542

\bibitem{} Hildebrand R.H., 1983, QJRAS 24, 267

\bibitem{} Knapp G.R., Guhathakurta P., Kim D.-W., Jura M., 1989,  ApJ 70,
329

\bibitem{} Knapp G.R., Gunn J.E., Wynn-Williams C.G., 1992, ApJ 399, 76

\bibitem{} Kwan J., Xie S., 1992, ApJ 398, 105

\bibitem{} Mathis J.S., Rumpl W., Nordsiek K.H., 1977, ApJ 217, 425

\bibitem{} Roberts M.S., Hogg D.E., Bregman J.N., Forman W.R., Jones C.,
1991, ApJS 75, 751

\bibitem{} Sandage A., Tammann G.A., 1981, ``A Revised Shapley-Ames Catalog
of Bright Galaxies'', (Carnegie Institution of 
Washington Publication: Washington) (RSA)

\bibitem{} Sparks W.B., Macchetto F., Golombek D., 1989, ApJ 345, 153

\bibitem{} Thronson H.A., Jr, Telesco C.M., 1986, ApJ 311, 98

\bibitem{} Thuan T.X., Sauvage M., 1992, A\&AS 92, 749

\bibitem{} Tsai J.C., Mathews W.G., 1995, ApJ 448, 84

\bibitem{} Tsai J.C., Mathews W.G., 1996, ApJ 468, 571

\bibitem{} V\'eron M.-P., V\'eron P., 1988, A\&A 204, 28

\bibitem{} Witt A.N, Thronson H.A. Jr., Capuano J.M. Jr, 1992, ApJ 393, 611

\bibitem{} Young J.S., Xie S., Kenney J.D., Rice W.L., 1989, ApJS 70, 699
\end{thebibliography}
\end{document}